\documentclass{article}
\usepackage{spconf,amsmath,graphicx,hyperref,booktabs,comment,subcaption}
\usepackage[normalem]{ulem}
\usepackage[square,numbers]{natbib}

\title{Compact Neural TTS Voices for Accessibility}
\name{\begin{tabular}{c}Kunal Jain, Eoin Murphy, Deepanshu Gupta, Jonathan Dyke, Saumya Shah \\
\textit {Vasilieios Tsiaras, Petko Petkov, Alistair Conkie}\end{tabular}}

\address{Apple Inc.}
\begin{document}
\maketitle
\begin{abstract}
Contemporary text-to-speech solutions for accessibility applications can typically be classified into two categories: (i) device-based statistical parametric speech synthesis (SPSS) or unit selection (USEL) and (ii) cloud-based neural TTS.
SPSS and USEL offer low latency and low disk footprint at the expense of naturalness and audio quality. Cloud-based neural TTS systems provide significantly better audio quality and naturalness but regress in terms of latency and responsiveness, rendering these impractical for real-world applications. More recently, neural TTS models were made deployable to run on handheld devices. Nevertheless, latency remains higher than SPSS and USEL, while disk footprint prohibits pre-installation for multiple voices at once. 
In this work, we describe a high-quality compact neural TTS system achieving latency on the order of 15 ms with low disk footprint. The proposed solution is capable of running on low-power devices.

\end{abstract}
\section{Introduction}
\label{sec:intro}

In recent years, 
text-to-speech (TTS) systems have emerged as crucial tools for enhancing accessibility across various domains.
These systems convert written text into spoken words, 
providing essential support for individuals with visual impairments, 
reading disabilities, 
and other conditions that hinder text comprehension \cite{smartspecs}.
By enabling auditory consumption of information, 
TTS systems empower users to engage with digital content independently and effectively.
For instance, 
screen readers equipped with TTS capabilities facilitate access to computers and smartphones for users with visual impairments, 
offering them the ability to navigate interfaces, 
read documents, 
and interact with online resources \cite{screen_reader}.
In educational settings, 
TTS can aid students with dyslexia, aphasia etc. by providing an alternative means to process and understand written material \cite{dyslexia_tts}.

For a TTS system used in accessibility settings, other than the need to sound more natural  \cite{need_for_sounding_natural}, additional performance considerations should be taken into account. Stricter latency requirements in these settings is motivated by {\cite{dang2024livespeechlowlatencyzeroshottexttospeech}}. For deployment in resource constrained environments, lower disk footprint and FLOPs also become important.

Traditional TTS systems used statistical approaches like Hidden Markov Models \cite{SPS2} for modeling text to acoustic features. Database-driven unit selection methods to generate audios made up of concatenated speech units have also been widely used \cite{US1}.
These systems lack naturalness and sound very robotic, which can be a sub-optimal experience for individuals.
Hence, 
there has been a surge in the research and development of systems that sound very natural and produce almost a human-like speech.
These recent approaches \cite{tan2021surveyneuralspeechsynthesis} are mostly deep learning based with large runtime computation and disk footprint requirement, which can be a hindrance to deployment on a low-powered device.
We develop a compact neural network based TTS system that is not only resource efficient (occupies only about 18 MB) on device, but also computationally efficient in comparison to neural TTS systems typically used on edge devices. 
In the next section, 
we outline the baseline architecture used in typical neural TTS systems.
These systems use high resources and are not directly usable as a system for accessibility purposes in a resource constrained setting.
In the successive sections, 
we discuss the optimizations necessary to compress the system's neural components, while meeting various considerations discussed earlier, for application in an accessibility setting.

\begin{figure*}
    \centering
    \begin{subfigure}{0.31\textwidth}
        \centering
        \includegraphics[width=0.65\linewidth, height=5cm]{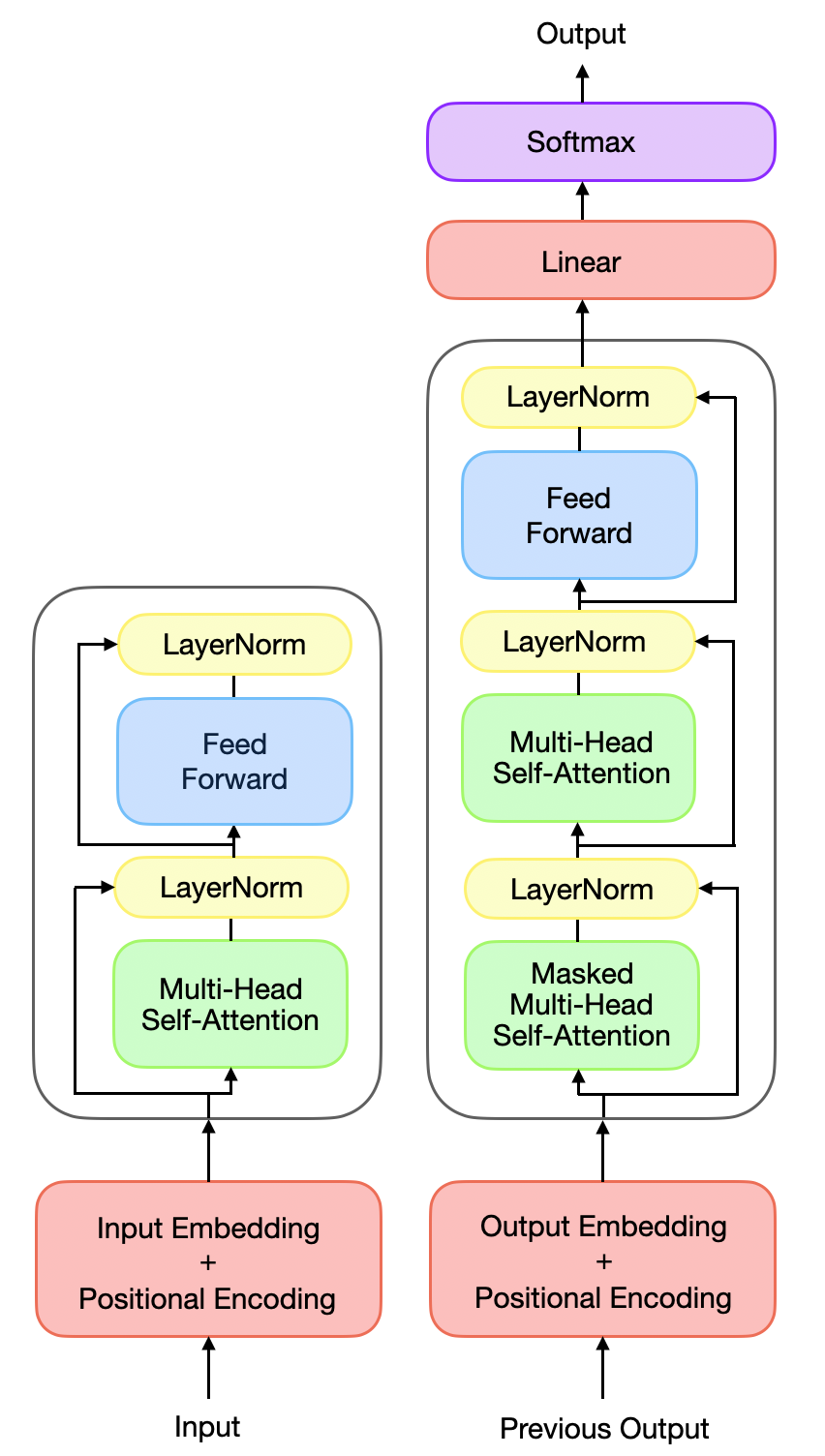}
        \caption{Transformer based Frontend}
        \label{fig:parta}
    \end{subfigure}
    \hspace{0.02\textwidth}
    \begin{subfigure}{0.31\textwidth}
        \centering
        \includegraphics[width=0.8\linewidth, height=5cm]{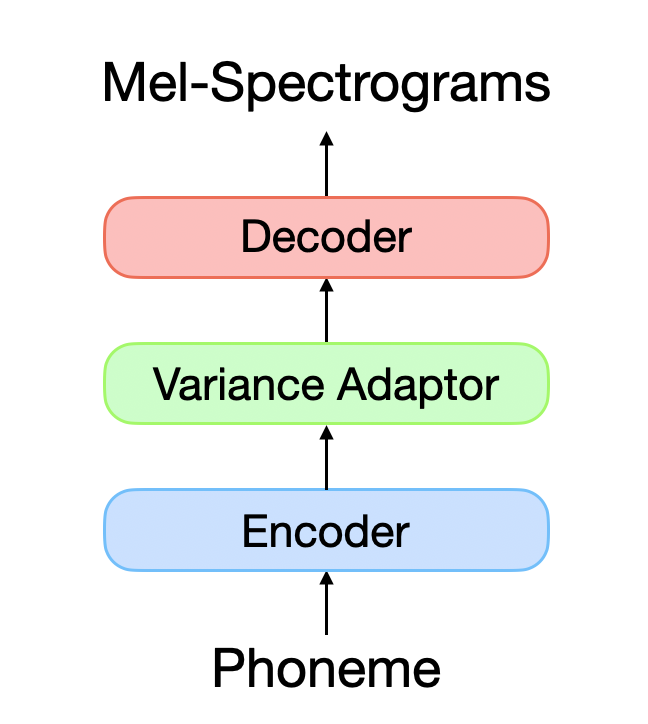}
        \caption{FastSpeech2 based Acoustic Model}
        \label{fig:partb}
    \end{subfigure}
    \hspace{0.02\textwidth}
    \begin{subfigure}{0.31\textwidth}
        \includegraphics[width=\linewidth, height=5cm]{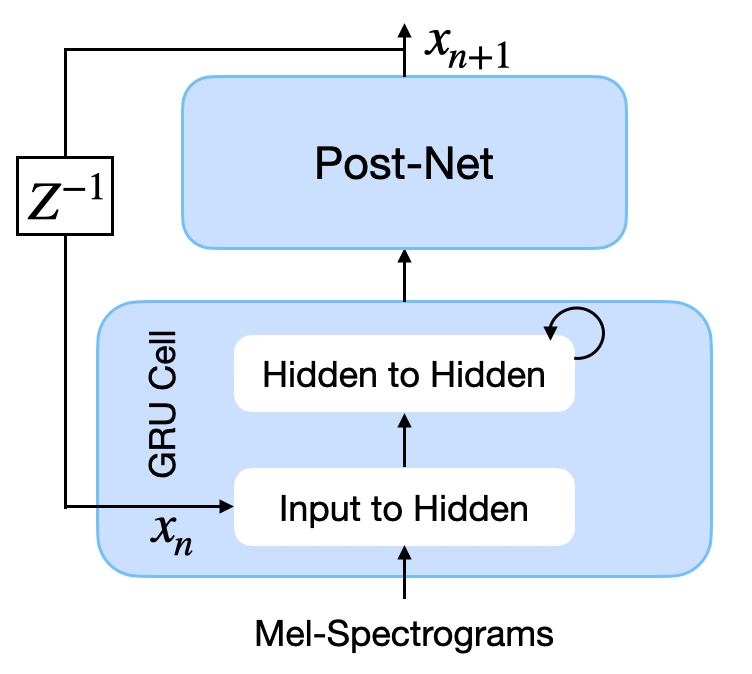}
        \caption{WaveRNN based Vocoder}
        \label{fig:partc}
    \end{subfigure}
    \caption{Neural Components in the Baseline Architecture}
    \label{fig:baselinee2e}
\end{figure*}

In this section, 
we describe an industry standard neural Text-To-Speech system and for the purpose of comparison, we call this our baseline system.  
The baseline generates audio given an unnormalized text sequence.
The system consists of three neural components placed sequentially: a text processing frontend, 
an acoustic model and a vocoder, illustrated in Figure \ref{fig:baselinee2e}. 
Each of these components is conventionally trained independently.

\subsection{Text Processing Frontend}
\label{subsec:example}
The frontend (FE) is the first stage of speech synthesis, and
is responsible for text tokenization, 
verbalization of non-standard words \cite{Sproat2001NormalizationON}, 
grapheme-to-phoneme conversion \cite{bisani2008joint}, 
homograph disambiguation \cite{yarowsky1997homograph},
and post-lexical analysis \cite{black2000building}. 
The most important aspect of the FE is Grapheme-to-Phoneme (G2P) conversion.
G2P converts input text from a language specific orthographic system to a 
language agnostic phonemic sequence.
To simplify and unify all of the various processing steps in FE, 
a Transformer based encoder-decoder architecture, as illustrated in Figure \ref{fig:parta}, is used for the baseline's frontend.
The autoregressive phoneme sequence prediction strategy is described in \cite{conkie2020scalablemultilingualfrontendtts}.
Traditionally, G2P modeling has been done at the word level, however,
modeling at the sentence level enables contextual disambiguation. 
This modification improves performance on ambiguities in the input text, for example with 
homograph disambiguation.


\subsection{Acoustic Model}
\label{subsec:accoustic}
The acoustic model in a text-to-speech system converts the phonemes predicted by the frontend into acoustic representations, typically in the form of mel-spectrograms.
FastSpeech2 (FS2) \cite{fastspeech2} is a popular acoustic model typically used for this stage. In addition to predicting mel spectrograms, FS2 also learns to model acoustic features like duration, pitch and energy which help in predicting more contextually accurate spectrograms. 
This model's ability to generate high-quality speech with fast and stable training has made it a popular choice for low-latency applications.
We use FS2 as our baseline acoustic model for the purpose of this work as shown in Figure \ref{fig:partb}.

\subsection{Vocoder}
\label{subsec:vocoder}
The vocoder is tasked with transforming these acoustic representations into raw waveforms.
WaveRNN, illustrated in Figure \ref{fig:partc}, a powerful neural vocoder, achieves high-fidelity speech synthesis by using a recurrent neural network architecture optimized for efficiency without sacrificing quality \cite{kalchbrenner2018efficient}. 
WaveRNN can produce natural-sounding speech in real-time, making it suitable for deployment in resource-constrained environments.

\section{Optimizations}
\label{sec:format}

In this section, 
we focus on optimizations of each neural component that enabled us to successfully employ the overall system on low-power devices under strict resource and latency constraints for use in accessibility settings.

\subsection{Text Processing Front End}
\label{subsec:example}

\begin{figure}
    \centering
    \includegraphics[width=0.5\textwidth]{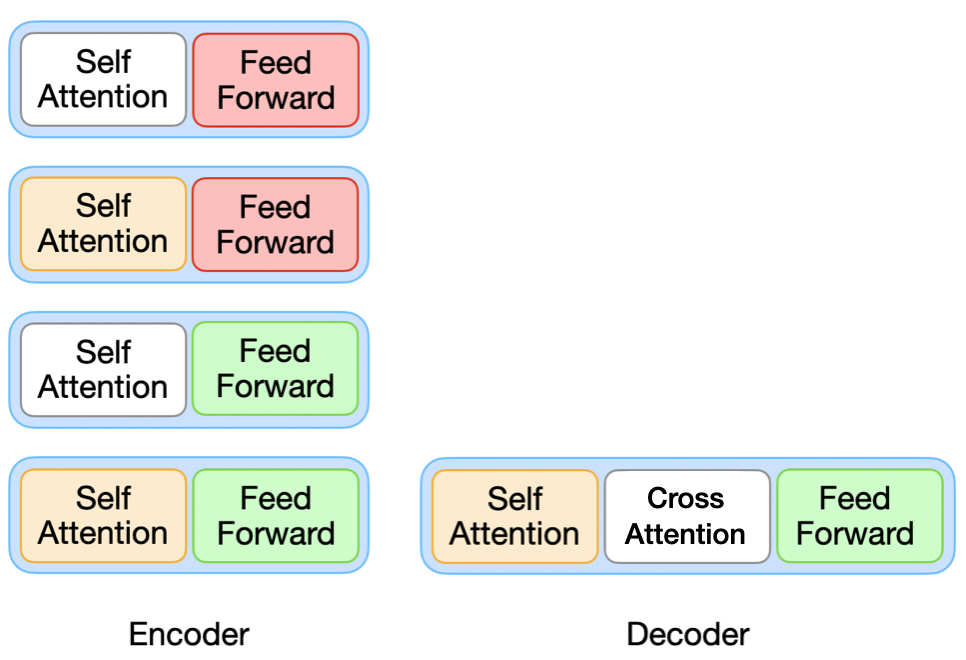} 
    \caption{Frontend parameter sharing. The same color indicates same parameters. The left are the layers of encoder and right is the only layer of decoder in the optimized frontend. As shown, we do sharing across encoder layers as well as across encoder and decoder for attention weights. Bias terms are independent for layer specific adaptation}
    \label{fig:frontend_optimization}
\end{figure}

We used a combination of techniques to achieve the optimization of the baseline frontend without major regression in performance. Inspired by studies on Machine Translation tasks, we used a deep encoder and shallow decoder setting \cite{deepencodershallowdecoder}. Such an encoder-favored parameterization ensures that the features learned by the encoder, which are later used in the decoder's cross attention, generalize better and retain sentence level information well. Further, having a smaller decoder which is used multiple times during autoregressive inference is desirable. In addition, redundant computations in the autoregressive decoding are reduced using key-value caching.
We used parameter sharing motivated by studies on Transformers \cite{lessonsparametersharinglayers} to reduce the frontend's disk footprint. In particular, we used strategies suggested in the paper \cite{edgeformerparameterefficient}. Interleaved sharing of attention weights between encoder and decoder, sharing of feed forward network weights among the encoder layers and load balanced parameterization across the encoder and decoder worked well when applied to baseline frontend. With such scale of sharing involved, we allowed the bias terms to be independent across layers for possible layer adaptation as discussed in paper \cite{bitfit}. Figure \ref{fig:frontend_optimization} illustrates the parameter sharing.

\subsection{Acoustic Model}
\label{subsec:example}

Post-training quantization from FP16 to INT8 \cite{Zhao2020Linear} was applied to the baseline FastSpeech2-based acoustic model to reduce the weights' footprint and improve computation speed. We further experimented both with smaller and a decreased number of convolutional filters. This allowed us to massively reduce the acoustic model size, as discussed in experiments section.

\subsection{Vocoder}
\label{subsec:example}
\begin{figure}
    \centering
    \includegraphics[width=0.38\textwidth]{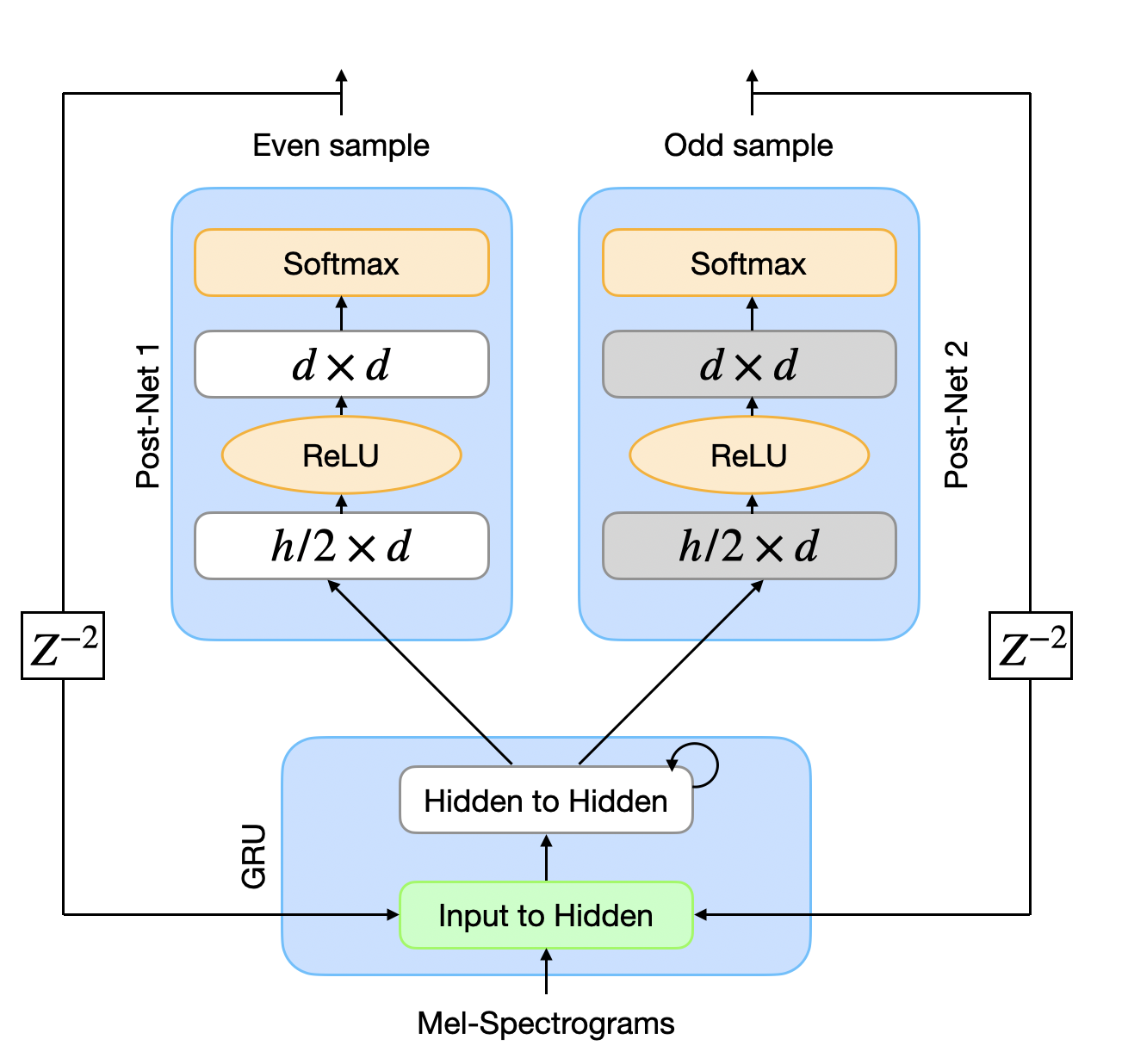} 
    \caption{Subscale WaveRNN produces two samples per iteration. The hidden-to-hidden matrix and the linear layers of the post-nets are sparse matrices, while the input-to-hidden matrix is dense. The hidden dimension is denoted by $h$, while the output dimension is denoted by $d$. Samples are produced by sampling the categorical distribution of the corresponding softmax.}
    \label{fig:subscale}
\end{figure}
WaveRNN is the slowest component of the system because it is called for every generated audio sample. A Real Time Factor (RTF) above 1.5x is required for WaveRNN to stream synthesized speech in real time while accommodating all additional processing, including the front-end and FS2. This is achievable on recent mobile devices \cite{Achanta2021}, and can also be achieved on low-power devices by using a) subscale generation of two samples per iteration and b) sparsity in the hidden-to-hidden mapping weights and in the fully-connected layers in the post-nets of the vocoder \cite{kalchbrenner2018efficient} (see Figure \ref{fig:subscale}). Sparsity is progressively learned during training.

\section{Experiments}
\label{sec:format}

\subsection{Datasets}
Our FE training datasets contain 5 million examples scraped from Wikipedia 
and transcribed using our production FE system. 
Datasets for training a FE can be created using \cite{mortensen2018epitran} and \cite{Sproat2001NormalizationON}. 

To train our acoustic model and vocoder, we use a proprietary dataset comprised of approximately 36 hours of recordings of a US English female speaker. Additional speech processing on recordings yields the ground truth acoustic features which are used for training the acoustic model and vocoder.
Some open source alternatives of parallel text and speech include LibriSpeech \cite{zen2019librittscorpusderivedlibrispeech}.

\subsection{Results}

Table \ref{tab:overall_metrics} shows the comparison between the baseline and our proposed optimized system. We performed a Mean Opinion Score (MOS) test with 45 native English listeners grading 1500 sentences predicted by both these models. For reference, the actual human recordings were rated 4.51 in the same environment. We observe a 4x reduction in footprint with very minimal MOS regression ($<3\%$). Other than the reduction in parameters, the reduced compute precision also play an important role in the improved latency we observe in the table. For all latency computations in the paper, we took the average across several sentences processed by the TTS system on a recent iOS device. For a detailed quality comparison, we also show MOS scores for the two systems across multiple domains: dialogues, navigation \& answers retrieved from knowledge graph, in Table \ref{tab:domain_wise_breakdown}. The results from this table show the compact model's quality remaining consistent across domains.

\begin{table}
    \centering
    \begin{tabular}{|c|c|c|}
        \toprule
        \textbf{Parameter} & \textbf{Baseline} & \textbf{Optimized} \\
        \midrule
        \hline
        Footprint & 73MB & 18MB \\
        \hline
        MOS & 4.19 & 4.09 \\
        \hline
        Latency & 27ms & 13ms \\
        \hline
        \bottomrule
    \end{tabular}
    \caption{Baseline vs Overall Optimized Metrics} 
    \label{tab:overall_metrics}
\end{table}

\begin{table}
    \centering
    \begin{tabular}{|c|c|c|}
        \toprule
        \textbf{Domain} & \textbf{Baseline} & \textbf{Optimized} \\
        \midrule
        \hline
        Dialogues & 4.29 & 4.13 \\
        \hline
        Navigation & 4.21 & 4.08 \\
        \hline
        Knowledge Graph & 4.06 & 3.91 \\
        \hline
        Overall & 4.19 & 4.09 \\
        \hline
        \bottomrule
    \end{tabular}
    \caption{MOS Across Domains} 
    \label{tab:domain_wise_breakdown}
\end{table}

\newpage
We conducted an ablation study, applying optimizations individually to each component and comparing metrics to the baseline. This ablation clarified each optimization's contribution to improvements in disk footprint and latency. Table \ref{tab:frontend_optimized_metrics} shows the comparison between the baseline and the system with only frontend optimizations applied. The `FE Config' in this table indicates the number of encoder and decoder layers used. 20-1(20 encoder layers and 1 decoder layer) is desirable for faster autoregressive decoding.
It's worth highlighting that the MOS test for this was carried out in a setting different from fully optimized \& baseline model and hence, isn't directly comparable even though it scored above 4.0. This holds true for Tables \ref{tab:acoustic_optimized_metrics} and  \ref{tab:vocoder_optimized_metrics}, which show comparisons for acoustic optimized and vocoder optimized respectively with baseline.

\begin{table}
    \centering
    \begin{tabular}{|c|c|c|}
        \toprule
        \textbf{Parameter} & \textbf{Baseline} & \textbf{FE Optimized} \\
        \midrule
        \hline
        FE Footprint & 28MB & 12MB \\
        \hline
        Total Footprint & 73MB & 57MB \\
        \hline
        FE Latency & 12ms & 7ms \\
        \hline
        Total Latency & 27ms & 13ms \\
        \hline
        FE Config. & 16-5 & 20-1 \\
        \hline
        \bottomrule
    \end{tabular}
    \caption{Frontend Optimized Metrics} 
    \label{tab:frontend_optimized_metrics}
\end{table}

\begin{table}
    \centering
    \begin{tabular}{|c|c|c|}
        \toprule
        \textbf{Parameter} & \textbf{Baseline} & \textbf{Acoustic Optimized} \\
        \midrule
        \hline
        Acoustic Footprint & 34MB & 2.6MB \\
        \hline
        Total Footprint & 73MB & 42MB \\
        \hline
        Acoustic Latency & 3ms & 1ms \\
        \hline
        Total Latency & 27ms & 13ms \\
        \hline
        \bottomrule
    \end{tabular}
    \caption{Acoustic Model Optimized Metrics} 
    \label{tab:acoustic_optimized_metrics}
\end{table}

\begin{table}
    \centering
    \begin{tabular}{|c|c|c|}
        \toprule
        \textbf{Parameter} & \textbf{Baseline} & \textbf{Vocoder Optimized} \\
        \midrule
        \hline
        Vocoder Footprint & 10.5MB & 3.1MB \\
        \hline
        Total Footprint & 73MB & 66MB \\
        \hline
        Vocoder Latency & 12ms & 5ms \\
        \hline
        Total Latency & 27ms & 13ms \\
        \hline
        \bottomrule
    \end{tabular}
    \caption{Vocoder Optimized Metrics} 
    \label{tab:vocoder_optimized_metrics}
\end{table}

If we want to deploy more voices in our TTS system for the same language, we can leverage the same frontend and we would only need to train acoustic and vocoder model from the desired voice's recordings. From Tables \ref{tab:frontend_optimized_metrics}, \ref{tab:acoustic_optimized_metrics} and \ref{tab:vocoder_optimized_metrics}, we observe that the frontend has most of the footprint post-optimization among the three components. This means we can scale additional voices with very small extra footprint (~5.7MB per additional voice taking additional optimized acoustic and vocoder into account). Thus, by leveraging a multilingual frontend, we can scale our system to multiple languages and voices by only an additional 5MB footprint per add-on. The work on a unified compact multilingual frontend is left as a potential future work.

\section{Conclusion}
\label{sec:conclusion}

In this paper, we use the recent neural advances in TTS synthesis systems to deliver high quality, low latency and super-small TTS system for 
special need individuals and accessibility use cases. 
Our effective use of optimizations, like Quantization, Weight Sharing, Sparsity, KV-Caching etc. allows us to remove redundant parameters and store them with a smaller device footprint.  
These optimizations allow us to achieve a high level of model compression and low inference latency, 
a core requirement for accessibility usage. 
As far as we are aware, our system is the first system to occupy less than 20MB disk space, and with a latency on the order of 15ms, whilst delivering an MOS score consistently close to a high quality production system. 

\section{Future Work}
\label{sec:future}
A multilingual compact frontend which allows for scaling TTS systems quickly for different languages and voices is an exciting research direction. 
One key area that could lead to further optimization and naturalness improvements is having a single 
end-to-end speech synthesis system instead of cascaded models. In the current setup, an erroneous prediction by an upstream model can directly impact the final output.
As technology progresses, we believe a future endeavor could create compact single model TTS solutions that cater to multiple languages, 
which we leave for future work.

\vfill
\pagebreak

\bibliographystyle{IEEEtran}
\setlength{\bibsep}{0pt plus 0.3ex}
\bibliography{refs}

\begin{thebibliography}{10}
\providecommand{\url}[1]{#1}
\csname url@samestyle\endcsname
\providecommand{\newblock}{\relax}
\providecommand{\bibinfo}[2]{#2}
\providecommand{\BIBentrySTDinterwordspacing}{\spaceskip=0pt\relax}
\providecommand{\BIBentryALTinterwordstretchfactor}{4}
\providecommand{\BIBentryALTinterwordspacing}{\spaceskip=\fontdimen2\font plus
\BIBentryALTinterwordstretchfactor\fontdimen3\font minus
  \fontdimen4\font\relax}
\providecommand{\BIBforeignlanguage}[2]{{%
\expandafter\ifx\csname l@#1\endcsname\relax
\typeout{** WARNING: IEEEtran.bst: No hyphenation pattern has been}%
\typeout{** loaded for the language `#1'. Using the pattern for}%
\typeout{** the default language instead.}%
\else
\language=\csname l@#1\endcsname
\fi
#2}}
\providecommand{\BIBdecl}{\relax}
\BIBdecl

\bibitem{smartspecs}
R.~Ani, E.~Maria, J.~J. Joyce, V.~Sakkaravarthy, and M.~A. Raja, ``Smart
  {S}pecs: {V}oice assisted text reading system for visually impaired persons
  using {TTS} method,'' in \emph{2017 International Conference on Innovations
  in Green Energy and Healthcare Technologies (IGEHT)}, 2017, pp. 1--6.

\bibitem{screen_reader}
A.~Chalamandaris, S.~Karabetsos, P.~Tsiakoulis, and S.~Raptis, ``A unit
  selection text-to-speech synthesis system optimized for use with screen
  readers,'' \emph{IEEE Transactions on Consumer Electronics}, vol.~56, no.~3,
  pp. 1890--1897, 2010.

\bibitem{dyslexia_tts}
\BIBentryALTinterwordspacing
V.~Giannouli and M.~Banou, ``The intelligibility and comprehension of synthetic
  versus natural speech in dyslexic students,'' \emph{Disability and
  Rehabilitation: Assistive Technology}, vol.~15, no.~8, pp. 898--907, 2020,
  pMID: 31339391. [Online]. Available:
  \url{https://doi.org/10.1080/17483107.2019.1629111}
\BIBentrySTDinterwordspacing

\bibitem{need_for_sounding_natural}
\BIBentryALTinterwordspacing
G.~Cistola, A.~Peiró-Lilja, G.~Cámbara, I.~van~der Meulen, and M.~Farrús,
  ``Influence of {TTS} {S}ystems {P}erformance on {R}eaction {T}imes in
  {P}eople with {A}phasia,'' \emph{Applied Sciences}, vol.~11, no.~23, 2021.
  [Online]. Available: \url{https://www.mdpi.com/2076-3417/11/23/11320}
\BIBentrySTDinterwordspacing

\bibitem{dang2024livespeechlowlatencyzeroshottexttospeech}
\BIBentryALTinterwordspacing
T.~Dang, D.~Aponte, D.~Tran, and K.~Koishida, ``Livespeech: Low-latency
  zero-shot text-to-speech via autoregressive modeling of audio discrete
  codes,'' 2024. [Online]. Available: \url{https://arxiv.org/abs/2406.02897}
\BIBentrySTDinterwordspacing

\bibitem{SPS2}
\BIBentryALTinterwordspacing
K.~Tokuda, Y.~Nankaku, T.~Toda, H.~Zen, J.~Yamagishi, and K.~Oura, ``Speech
  {S}ynthesis {B}ased on {H}idden {M}arkov {M}odels,'' \emph{Proceedings of the
  IEEE}, vol. 101, pp. 1234--1252, 2013. [Online]. Available:
  \url{https://api.semanticscholar.org/CorpusID:33895269}
\BIBentrySTDinterwordspacing

\bibitem{US1}
\BIBentryALTinterwordspacing
A.~J. Hunt and A.~W. Black, ``Unit selection in a concatenative speech
  synthesis system using a large speech database,'' \emph{1996 IEEE
  International Conference on Acoustics, Speech, and Signal Processing
  Conference Proceedings}, vol.~1, pp. 373--376 vol. 1, 1996. [Online].
  Available: \url{https://api.semanticscholar.org/CorpusID:14621185}
\BIBentrySTDinterwordspacing

\bibitem{tan2021surveyneuralspeechsynthesis}
\BIBentryALTinterwordspacing
X.~Tan, T.~Qin, F.~Soong, and T.-Y. Liu, ``A survey on neural speech
  synthesis,'' 2021. [Online]. Available:
  \url{https://arxiv.org/abs/2106.15561}
\BIBentrySTDinterwordspacing

\bibitem{Sproat2001NormalizationON}
\BIBentryALTinterwordspacing
R.~Sproat, A.~W. Black, S.~F. Chen, S.~Kumar, M.~Ostendorf, and C.~D. Richards,
  ``Normalization of non-standard words,'' \emph{Comput. Speech Lang.},
  vol.~15, pp. 287--333, 2001. [Online]. Available:
  \url{https://api.semanticscholar.org/CorpusID:16861729}
\BIBentrySTDinterwordspacing

\bibitem{bisani2008joint}
M.~Bisani and H.~Ney, ``Joint-sequence models for grapheme-to-phoneme
  conversion,'' \emph{Speech communication}, vol.~50, no.~5, pp. 434--451,
  2008.

\bibitem{yarowsky1997homograph}
D.~Yarowsky, ``Homograph disambiguation in text-to-speech synthesis,'' in
  \emph{Progress in speech synthesis}.\hskip 1em plus 0.5em minus 0.4em\relax
  Springer, 1997, pp. 157--172.

\bibitem{black2000building}
A.~Black and K.~Lenzo, ``Building voices in the {F}estival speech synthesis
  system,'' 2000.

\bibitem{conkie2020scalablemultilingualfrontendtts}
\BIBentryALTinterwordspacing
A.~Conkie and A.~Finch, ``Scalable {M}ultilingual {F}rontend for {TTS},'' 2020.
  [Online]. Available: \url{https://arxiv.org/abs/2004.04934}
\BIBentrySTDinterwordspacing

\bibitem{fastspeech2}
\BIBentryALTinterwordspacing
Y.~Ren, C.~Hu, X.~Tan, T.~Qin, S.~Zhao, Z.~Zhao, and T.-Y. Liu, ``Fast{S}peech
  2: {F}ast and {H}igh-{Q}uality {E}nd-to-{E}nd {T}ext to {S}peech,'' 2022.
  [Online]. Available: \url{https://arxiv.org/abs/2006.04558}
\BIBentrySTDinterwordspacing

\bibitem{kalchbrenner2018efficient}
N.~Kalchbrenner, E.~Elsen, K.~Simonyan, S.~Noury, N.~Casagrande, E.~Lockhart,
  F.~Stimberg, A.~Oord, S.~Dieleman, and K.~Kavukcuoglu, ``Efficient neural
  audio synthesis,'' in \emph{International Conference on Machine
  Learning}.\hskip 1em plus 0.5em minus 0.4em\relax PMLR, 2018, pp. 2410--2419.

\bibitem{deepencodershallowdecoder}
\BIBentryALTinterwordspacing
J.~Kasai, N.~Pappas, H.~Peng, J.~Cross, and N.~A. Smith, ``Deep {E}ncoder,
  {S}hallow {D}ecoder: {R}eevaluating {N}on-autoregressive {M}achine
  {T}ranslation,'' 2021. [Online]. Available:
  \url{https://arxiv.org/abs/2006.10369}
\BIBentrySTDinterwordspacing

\bibitem{lessonsparametersharinglayers}
\BIBentryALTinterwordspacing
S.~Takase and S.~Kiyono, ``Lessons on {P}arameter {S}haring across {L}ayers in
  {T}ransformers,'' 2023. [Online]. Available:
  \url{https://arxiv.org/abs/2104.06022}
\BIBentrySTDinterwordspacing

\bibitem{edgeformerparameterefficient}
\BIBentryALTinterwordspacing
T.~Ge, S.-Q. Chen, and F.~Wei, ``Edge{F}ormer: A {P}arameter-{E}fficient
  {T}ransformer for {O}n-{D}evice {S}eq2seq {G}eneration,'' 2022. [Online].
  Available: \url{https://arxiv.org/abs/2202.07959}
\BIBentrySTDinterwordspacing

\bibitem{bitfit}
\BIBentryALTinterwordspacing
E.~B. Zaken, S.~Ravfogel, and Y.~Goldberg, ``Bit{F}it: Simple
  {P}arameter-efficient {F}ine-tuning for {T}ransformer-based {M}asked
  {L}anguage-models,'' 2022. [Online]. Available:
  \url{https://arxiv.org/abs/2106.10199}
\BIBentrySTDinterwordspacing

\bibitem{Zhao2020Linear}
\BIBentryALTinterwordspacing
X.~Zhao, Y.~Wang, X.~Cai, C.~Liu, and L.~Zhang, ``Linear symmetric quantization
  of neural networks for low-precision integer hardware,'' in
  \emph{International Conference on Learning Representations}, 2020. [Online].
  Available: \url{https://openreview.net/forum?id=H1lBj2VFPS}
\BIBentrySTDinterwordspacing

\bibitem{Achanta2021}
S.~Achanta, A.~Antony, L.~Golipour, J.~Li, T.~Raitio, R.~Rasipuram, F.~Rossi,
  J.~Shi, J.~Upadhyay, D.~Winarsky, and H.~Zhang, ``On-device {N}eural {S}peech
  {S}ynthesis,'' in \emph{ASRU}, 2021, pp. 1155--1161.

\bibitem{mortensen2018epitran}
D.~R. Mortensen, S.~Dalmia, and P.~Littell, ``Epitran: {P}recision {G2P} for
  many languages,'' in \emph{Proceedings of the Eleventh International
  Conference on Language Resources and Evaluation (LREC 2018)}, 2018.

\bibitem{zen2019librittscorpusderivedlibrispeech}
\BIBentryALTinterwordspacing
H.~Zen, V.~Dang, R.~Clark, Y.~Zhang, R.~J. Weiss, Y.~Jia, Z.~Chen, and Y.~Wu,
  ``Libri{TTS}: A {C}orpus {D}erived from {L}ibri{S}peech for
  {T}ext-to-{S}peech,'' 2019. [Online]. Available:
  \url{https://arxiv.org/abs/1904.02882}
\BIBentrySTDinterwordspacing

\end{thebibliography}

\end{document}